\documentclass[12pt]{article}

\usepackage{epsf}
\usepackage{latexsym,amssymb,euscript}
\usepackage[dvips]{graphicx}
\usepackage{amsmath}

\begin{document}

\begin{center}
{\Large \bf Deconfinement and universality in the $3D$ $U(1)$ \newline
lattice gauge theory at finite temperature: \newline
study in the dual formulation}
\end{center}

\vskip 0.3cm
\centerline{O.~Borisenko$^{1\dagger}$, V.~Chelnokov$^{1*}$, 
M.~Gravina$^{2\ddagger}$, A.~Papa$^{2\P}$}

\vskip 0.6cm

\centerline{${}^1$ \sl Bogolyubov Institute for Theoretical Physics,}
\centerline{\sl National Academy of Sciences of Ukraine,}
\centerline{\sl UA-03680 Kiev, Ukraine}

\vskip 0.2cm

\centerline{${}^2$ \sl Dipartimento di Fisica, Universit\`a della 
Calabria,}
\centerline{\sl and Istituto Nazionale di Fisica Nucleare, 
Gruppo collegato di Cosenza,}
\centerline{\sl I-87036 Arcavacata di Rende, Cosenza, Italy}

\vskip 0.6cm

\begin{abstract}
We study analytically and numerically the three-dimensional $U(1)$ lattice 
gauge theory at finite temperature in the dual formulation. For an appropriate 
disorder operator, we obtain the renormalization group equations describing the 
critical behavior of the model in the vicinity of the deconfinement phase 
transition. 
These equations are used to check the validity of the Svetitsky-Yaffe 
conjecture regarding the critical behavior of the lattice $U(1)$ model. 
Furthermore, we perform numerical simulations of the model for 
$N_t = 1, 2, 4, 8$ and compute, by a cluster algorithm, the dual correlation 
functions and the corresponding second moment correlation length. 
In this way we locate the position of the critical point and calculate
critical indices. 
\end{abstract}

\vfill
\hrule
\vspace{0.3cm}
{\it e-mail addresses}:
$^\dagger$oleg@bitp.kiev.ua, \  $^*$chelnokov@bitp.kiev.ua,
\  $^{\ddagger}$gravina@fis.unical.it, \ \ $^{\P}$papa@fis.unical.it 

\section{Introduction} 

The Svetitsky-Yaffe conjecture~\cite{svetitsky} states that, if the correlation 
length of a $D$-dimensional finite-temperature gauge theory with a given 
symmetry group $G$ diverges at the transition point, then the gauge theory 
belongs to the same universality class of the $(D-1)$-dimensional spin model 
possessing the center of the group $G$ as global symmetry. 
This connection at criticality is relevant for a large class of gauge systems 
and turns out to be fundamental in understanding the deconfinement phase 
transition of heavy colored charges, {\it i.e.} pure gauge QCD.

The validity of the Svetitsky-Yaffe conjecture in presence of an infinite-order 
phase transition is yet to be verified. This is the case, for instance, of 
the deconfinement phase transition in the finite-temperature $3D$ $U(1)$ 
lattice gauge theory (LGT). If the Svetitsky-Yaffe conjecture holds, 
this theory should belong to the universality class of the $2D$ $XY$ model, 
which is known to undergo a Berezinskii-Kosterlitz-Thouless 
(BKT)~\cite{berezin,kosterlitz1} phase transition. This means that the two 
models should exhibit the same critical indices, thus implying a common dynamical 
behavior at criticality. 
In general, critical indices are extracted from the study of the dependence
of the order parameter on the couplings of the theory near the transition
point. In finite-temperature $3D$ $U(1)$ LGT, as well as in the $2D$ $XY$ spin 
model, due to the Mermin-Wagner theorem~\cite{mwtheorem}, no spontaneous 
symmetry breaking can occur and, consequently, there exist no order parameter.
A critical index can nevertheless be defined in $3D$ $U(1)$ through the 
correlation function of two Polyakov loops, which is the counterpart of the
spin-spin correlator in the $2D$ $XY$ model. Indeed this correlation
function decreases with a power law for $\beta \geq \beta_{\rm c}$, thus implying a 
logarithmic potential between heavy electrons,
\begin{equation}
P(R) \ \asymp \ \frac{1}{R^{\eta (T)}} \ ,
\label{PLhight}
\end{equation}
$R\gg 1$ being the distance between test charges.
The critical index $\eta (T)$ is known from the renormalization group (RG) 
analysis of Ref.~\cite{kosterlitz1} and equals 1/4 at the critical
point of the BKT transition in $2D$ $XY$, $\eta (T_{\rm c}) =1/4$. 
For $\beta < \beta_{\rm c}$ one has instead
\begin{equation}
P(R) \ \asymp \ \exp \left [ -R/\xi (t)  \right ] \ ,
\label{PLlowt}
\end{equation}
where $t=\beta_{\rm c}/\beta -1$, the correlation length goes as 
$\xi \sim e^{bt^{-\nu}}$, and the critical index $\nu$ is equal to 1/2
in $2D$ $XY$. To determine the universality class of the finite-temperature 
$3D$ $U(1)$ LGT and to check if it coincides with that of $2D$ $XY$, the 
indices $\eta$ and $\nu$ must be calculated and compared with 1/4 and 1/2,
respectively~\footnote{If not otherwise specified, from now on $\eta$ 
will stand for $\eta(T_{\rm c})$.}.

On the analytical side, Parga studied the $3D$ $U(1)$ LGT in the Lagrangian 
formulation~\cite{parga}, explaining that at high temperatures the system 
becomes effectively two-dimensional and, in particular, the monopoles of the 
original $U(1)$ gauge theory become vortices of the $2D$ system. The partition 
function turns out to coincide (at the leading order of the high-temperature 
expansion) with that of the $2D$ $XY$ model in the Villain representation, 
thus supporting the Svetitsky-Yaffe conjecture. 
In the RG study of Refs.~\cite{svetitsky,borisenko,twist_rg}, 
high-temperature and dilute monopole gas approximations were used for the 
Villain formulation of finite-temperature $3D$ $U(1)$, which helped to 
derive a scaling law for the effective coupling between Polyakov loops with 
the lattice spacing. The resulting RG equations were shown to converge rapidly 
with the iteration number to the RG equations of the $2D$ $XY$ model. This 
represents a strong indication that, indeed, the nature of the phase 
transitions in the two models is the same. Moreover, since the scaling 
with the lattice spacing coincides in the two cases, one concludes that 
the critical index $\nu$ is the same in the two models.  
The critical points and the index $\eta$ in the Villain formulation of 
finite-temperature $3D$ $U(1)$ have been determined for various values
of $N_t$ in~\cite{twist_rg}, via numerical analysis of the RG equations,
confirming that $\eta=1/4$. 

On the numerical side, evidences that the deconfinement transition 
in the $3D$ $U(1)$ LGT belongs to the universality class of the $2D$ $XY$ 
model come from our study of the phase transitions in $3D$ $Z(N)$ 
LGT~\cite{3dzn_strcoupl,3dzn_isotropic1,3dzn_isotropic2,pos_lat} for large 
values of $N$. These models exhibit two separate BKT-like phase transitions
at critical couplings, say, $\beta_2(N)$ and $\beta_1(N)$, with 
$\beta_2(N)>\beta_1(N)$. While $\beta_2(N)$ diverges with $N$, the critical 
coupling $\beta_1(N)$ seems to converge to a value, which is expected, 
on general grounds, to represent the critical point of the $3D$ $U(1)$ LGT.
Moreover, it turns out the critical indices at the transition in $\beta_1(N)$
do not depend on $N$ in the interval $N\in [5-20]$ and are compatible
with the universality class of the $2D$ $XY$ model. The same phenomenon
occurs in the $2D$ $Z(N)$ spin models for large $N$. This strongly suggests that 
$3D$ $U(1)$ near criticality belongs to the universality class of the $2D$ 
$XY$ model. 

The first direct simulation of $3D$ $U(1)$ on the lattice was performed on 
$L^2\times N_t$ lattices with $L=16, 32$ and $N_t=4,6,8$ in~\cite{mcfinitet}. 
These authors confirmed the expected BKT nature of the phase transition in 
the gauge model, but reported a critical index almost three times larger than
that predicted for the $2D$ $XY$ model, $\eta \approx 0.78$.
The more recent analytical and numerical study of Ref.~\cite{beta_szero} 
indicated that, at least in the limit of vanishing spatial coupling, 
$\beta_s \to 0$, {\it i.e.} on an anisotropic lattice with decoupled 
space-like plaquettes, the $3D$ $U(1)$ LGT exhibits the same critical 
behavior of the $2D$ $XY$ spin model. Numerical simulations of the isotropic 
model, on lattices with $N_t=8$ and spatial extension up to $L=256$, revealed 
instead that $\eta\approx 0.49$, which is still far from the $XY$ 
value~\cite{u1_isotropic}, thus leaving open the question about the 
universality class of $3D$ $U(1)$ LGT with nonvanishing $\beta_s$.

A possible explanation for this mismatch at large $N_t$ can reside in the 
infinite order nature of the BKT transition. Indeed, the exponential divergence 
of the correlation length in a BKT transition implies a very slow, logarithmic 
convergence to the thermodynamic limit in the vicinity of the transition. As a 
consequence, very large volumes are required in order to enter the 
scaling region and safely extract critical indices.

In~\cite{3dzn_isotropic1,3dzn_isotropic2} we developed a strategy based on 
duality, according to which $Z(N)$ LGTs were mapped into spin models. This 
opened the way to the use of cluster algorithms, which facilitated the access 
to larger volumes, thus improving the description of criticality. In the 
present work we adjust that strategy to the study of the critical behavior of 
the $3D$ $U(1)$ LGT. This goes through the following steps:

\begin{itemize}
\item 
Calculate the disorder operator in the dual formulations of the $2D$ $XY$ 
spin model and in the $3D$ $U(1)$ LGT in the dilute gas approximation.

\item 
Derive and study the RG equations in the Villain formulation; compute the 
critical indices $\nu$ and $\eta$ and give an analytical prediction for the 
index $r$ related to the leading logarithmic correction.

This is the subject of Section~2. The main conclusion of these analytical 
investigations is that the critical behavior of the disorder operator of the 
$3D$ $U(1)$ LGT is governed by the critical behavior of the 
corresponding operator in the $2D$ $XY$ model. Therefore, the universality 
problem can be studied with the help of the disorder operator on the same 
theoretical footing as with the help of the Polyakov loop correlation function. 

\item 
Use the cluster algorithm in the dual formulation of $3D$ $U(1)$ to determine 
the second moment correlation length and to locate the critical points of the 
deconfinement transition; then, compute the critical indices from the large 
distance behavior of the disorder operator. 

This is done in Section~3. 

\end{itemize}

In Section~4 we summarize our results and draw our conclusions.

\section{Dual of $3D$ $U(1)$ LGT and disorder operator}

We work on a periodic $3D$ lattice $\Lambda = L^2\times N_t$ with spatial 
extension $L$ and temporal extension $N_t$. With the goal of performing a RG 
analysis, we introduce anisotropic dimensionless couplings as 
\begin{equation}
\beta_t = \frac{1}{g^2a_t}  \ , \;\;\;\;\; \beta_s = \frac{\xi}{g^2a_s} \ = \ 
\beta_t \ \xi^2 \ , \;\;\;\;\; \xi = \frac{a_t}{a_s} \ ,
\label{ancoupl}
\end{equation}
where $a_t$ ($a_s$) is lattice spacing in the time (space) direction, 
$g^2$ is the continuum coupling constant with dimension of an inverse length and
$\beta = a_tN_t$ is the inverse temperature. 
The dual of $3D$ $U(1)$ LGT is given by 
\begin{equation}
Z_{\Lambda}(\beta_n) \  = \ \sum_{\{ r(x) \} =-\infty}^{\infty} \  
\prod_x \ \prod_{n=1}^{3} \ I_{r(x)-r(x+e_n)}(\beta_n) \ . 
\label{Zdual_u1}
\end{equation}
Here $I_r(x)$ is the modified Bessel function of first kind and order $r$, 
$\beta_3=\beta_s$ and $\beta_1=\beta_2=\beta_t$.
When $\beta_s=0$ and $N_t=1$ the theory reduces to the dual of the $2D$ $XY$ 
model. 

The conventional disorder operator in the $XY$ model, 
\begin{equation} 
D(x,y) \ = \ \langle \ \exp \left [ i c ( r(x)-r(y) )   \right ] \ \rangle \ , 
\;\;\;\;\; \ 0 < c < 2\pi \;,
\label{disorder_xy}
\end{equation} 
defines the free energy of the vortex-antivortex pair. It obeys the following 
bound~\cite{xy_disorder}
\begin{equation} 
D(x,y) \  \leq \ | x - y |^{- c^2\gamma(\beta)} \ , \;\;\; 
\ \gamma(\beta) > 0 \ .
\label{disorder_bound}
\end{equation} 
A similar disorder operator in the $3D$ $U(1)$ LGT gives the free energy of 
a monopole-antimonopole pair. 
In this paper we use the following generalization of the disorder operator 
\begin{equation} 
D(x,y) \ = \ \langle \ \exp \left [ i \ \frac{c}{N_t} \ \sum_{x_3=0}^{N_t} 
( r(x)-r(y) )   \right ] \ \rangle \ .
\label{disorder_u1}
\end{equation} 
The reason for such definition will be explained shortly. 

We want to calculate the disorder operators~(\ref{disorder_xy}), 
(\ref{disorder_u1}) and derive RG equations from an effective coupling which 
describes the behavior of these operators near criticality. For simplicity 
we give details of the derivation for the $XY$ model and then explain how it 
can be extended to $U(1)$ LGT at finite temperature. When both couplings are 
large, it is customary to use the Villain approximation, {\it i.e.}
\begin{equation}
I_r(x)/I_0(x) \ \approx \ \exp \left ( - \frac{1}{2x} r^2  \right )  \ .
\label{PTVillainDef}
\end{equation} 
The Villain model, obtained by taking the approximation ({\ref{PTVillainDef}}),
is generally accepted to have the same universal properties as the original 
model~\cite{svetitsky,parga}.  

Let us consider first the general disorder operator 
\begin{equation} 
D(s) \ = \ \langle  \ 
\exp \left [ i \ \sum_x \  s(x) r(x) \right ] \ \rangle \ , 
\label{disorder_gen}
\end{equation} 
where $s(x)$ are sources for dual variables $r(x)$. 
Substituting~(\ref{PTVillainDef}) into the partition function~(\ref{Zdual_u1}),
we use the Poisson summation formula to perform the summation over the 
variables $r(x)$. In the case of the $XY$ model, the disorder 
operator~(\ref{disorder_gen}) is factorized in the product of the spin wave 
contribution and the contribution from the vortex 
configurations~($\beta_s=0$, $\beta_t=\beta$),
\begin{equation} 
 D(s) \ = \  D_{\rm sw}(s) \ D_{\rm v}(s) \ .
\label{PT_fact}
\end{equation} 
The spin wave contribution is given by (sum over repeated coordinates is 
understood)
\begin{equation}
D_{\rm sw}(s) \ = \ \exp \left[ -\frac{1}{4} \ \beta \  s(x)G_{x,y} s(y) \right] \ ,
\label{Dsw}
\end{equation} 
while the vortex part can be presented as 
\begin{equation} 
D_{\rm v}(s) \ =  \ D_{\rm v}^{-1}(s=0) \ 
\sum_{\{m_x\}} \exp \left[ -\pi^2 \beta \  m(x) G_{x,y} m(y) 
- \pi \beta m(x) G_{x,y} s(y) \right] \;.
\label{Dv}
\end{equation}
Here, $G_{x,y}\equiv G_{|x-y|}$ is the two-dimensional massless Green's function. 
Our following calculations are based on the dilute gas approximation, which can 
be used when $\beta$ is large enough. In this case the leading contribution 
comes from the configurations $m(x) = 0, \pm 1$ and, taking into account the 
neutrality of the vortex ensemble, we obtain after a long but standard algebra
\begin{equation}
D_{\rm v}(s) \ \approx \ \exp \left [ 2\pi^3\beta^2 
\int_1^{\infty}r^3 e^{-2\pi^2\beta D(r)} dr \ s(x)G_{x,y} s(y) \right ] \ ,
\label{zm_final}
\end{equation} 
where $D(r)=G_0-G_r$.
Taking the asymptotics of the $D(r)$ function and combining the last equation 
with Eq.~(\ref{Dsw}), we write down the result in the form 
\begin{equation}
D(s) \ = \ \exp \left[ -\frac{1}{4} \ \beta_{\rm eff} \  s(x)G_{x,y} s(y) \right] \ ,
\label{Dres}
\end{equation} 
where 
\begin{equation} 
\beta_{\rm eff} \ = \ \beta - 2\pi^3\beta^2 \ y^2 \ \int_1^{\infty} \ r^{3-2\pi\beta} 
\ dr \ ,
\label{beta_eff}
\end{equation}
and we have introduced the vortex activity as 
\begin{equation} 
y \ = \ 2  \ e^{-\frac{1}{2} \pi^2\beta} \ .
\label{vortex_act}
\end{equation}

The RG equations can be derived from the expression for $\beta_{\rm eff}$, 
by integrating in Eq.~(\ref{beta_eff}) between the length scales $a$ and 
$a+\delta a$, see {\it e.g}~\cite{nelson}. Denoting $t=\ln a$, one finds 
\begin{eqnarray}
\frac{d \beta}{d t} \ = \ - 2 \pi^3 y^2 \beta^2  \  , \ \;\;\;\;\;
\frac{d y}{d t} \ = \ y  \left(2 - \pi \beta \right) \ . 
\label{rg2}
\end{eqnarray} 
These equations coincide with the conventional RG equations for the $XY$ model. 
This should not come as a surprise, because $\beta_{\rm eff}$ derived above 
equals the corresponding effective coupling for the spin-spin correlation and 
for the twist free energy up to ${\cal{O}}(y^4)$~\cite{nelson}. 

To extend this result to the finite-temperature $U(1)$ LGT, we use the disorder 
operator~(\ref{disorder_u1}) and calculate it in the anisotropic 
model~(\ref{Zdual_u1}). The calculations follow closely the ones for 
the twist free energy~\cite{twist_rg} and lead to the same expression, again up 
to ${\cal{O}}(y^4)$, for the $\beta_{\rm eff}$ that controls the behavior of the 
twist (see formula~(24) in Ref.~\cite{twist_rg}). 
Hence, all the analysis of~\cite{twist_rg} remains valid for the disorder 
operator~(\ref{disorder_u1}). In particular, the fixed point of RG equations 
scales with $N_t$ as 
\begin{equation} 
\beta_t^{f} \ = \ \frac{2}{\pi} \ N_t \ . 
\label{beta_fixed}
\end{equation} 

An important consequence, relevant for our study, concerns the fall-off of the 
two-point disorder operator and the corresponding second-moment correlation 
length. Taking the sources in~(\ref{disorder_gen}) in the form 
\begin{equation} 
s(z) \ = \ \frac{c}{N_t} \ \left ( \delta_{z,x}-\delta_{z,y} \right ) \;,
\label{sources}
\end{equation}
we find the leading term to be  
\begin{equation} 
D(x,y) \ = \ \frac{\mbox{const}}{R^{\eta}} \ , \  \;\;\;\;\;
R \ = \ |x-y| \ . 
\label{disorder_falloff}
\end{equation}
The index $\eta$ for all $N_t$ is found to be 
\begin{equation} 
\eta(\beta_{\rm eff}) \ = \ \frac{c^2}{2\pi N_t} \ \beta_{\rm eff} \ . 
\label{eta_form}
\end{equation}
Since at the critical point $\beta_{\rm eff}$ takes the fixed point 
value~(\ref{beta_fixed}), we finally obtain the expression for the index 
$\eta$ at the phase transition point,  
\begin{equation} 
\eta \ = \ \left ( \frac{c}{\pi} \right )^2 \ . 
\label{eta_crit}
\end{equation}
In particular, it leads to $\eta=1/4$ when $c=\pi/2$, {\it i.e.} the value 
that equals the conventional value obtained from the spin-spin correlation 
function. The leading logarithmic correction to the power-like fall-off 
can also be easily computed at the critical point following the standard scheme 
(see, for instance, Section~4 of \cite{itzykson}),  
\begin{equation} 
D(x,y) \ = \ \frac{\mbox{const}}{R^{\eta_{\rm c}}} \ (\ln R)^{2r} \ , 
\label{disorder_falloff_log}
\end{equation} 
where 
\begin{equation} 
 r \ = \ - \left ( \frac{c}{2 \pi} \right )^2 \ . 
\label{index_r}
\end{equation} 
Note that, $r$ being negative, the leading logarithm appears in the 
denominator, in contrast to the spin-spin correlation function. 

The above observations imply that such RG-invariant quantities, like the 
second-moment correlation length and the Binder cumulant, take universal 
values that are independent of $N_t$ and are known for the $XY$ 
model~\cite{hasenbusch,binder_xy}. 

In the next Section we combine this observation with a cluster algorithm to 
locate critical points for different $N_t$ and to compute the index $\eta$. 

\section{Numerical data} 

In this Section we simulate the dual $3D$ $U(1)$ model on a $L^2 \times N_t$ 
lattice, using the cluster algorithm described in~\cite{dual_cluster}. 
We define the dual second-moment correlation length $\xi_2$ in the following 
way:
\begin{equation}
\xi_2  = \frac{\sqrt{\frac{\chi}{F}-1}}{2 \sin {\pi/L}} \; ,
\label{defxi2}
\end{equation}
\[
\chi  =  \left\langle \sum_{x,y} D(x,y) \right\rangle \; , \;\;\;\;\;
F  = \left\langle \sum_{x,y} e^{2\pi i (x_1 - y_1)/L} D(x,y) \right\rangle \; ,
\]
where $D(x,y)$ is the disorder operator defined in~(\ref{disorder_u1}) and 
$c$ in that equation is an arbitrary parameter,
defining the numerical value of $\xi_2$. 
In what follows we will consider $c = \pi/3, \pi/2$ and $2\pi/3$.

From the analytical expression for the correlation function, one can find the 
following scaling of $\xi_2$ with $L$ at the critical point, using the method 
described in~\cite{hasenbusch,binder_xy}:
\begin{equation}
\label{xi2scaling}
\frac{\xi_2}{L} = A_{\rm c} - \frac{B_{\rm c}}{\ln L + C}\; .
\end{equation}

The values $A_{\rm c}$ and $B_{\rm c}$ are calculated in spin wave approximation, taking 
$\beta_{\rm eff} = 2/\pi$. In that case 
\begin{eqnarray}
\chi(\beta) & = & \sum_{R} \exp \left( -\frac{c^2 \beta}{2} D(R) \right) \; , 
\nonumber \\
F(\beta)    & = & \sum_{R} \exp \left( -\frac{c^2 \beta}{2} D(R) \right) 
\cos \frac{2 \pi x}{L} \; , \nonumber \\
A_{\rm c} & = & \lim_{L \to \infty} \xi_2(\beta_{\rm eff}) \; , \nonumber \\
B_{\rm c} & = & \frac{1}{\pi} \lim_{L \to \infty} {\left. \frac{d \xi_2(\beta)}{d \beta} 
\right|}_{\beta=\beta_{\rm eff}} \; ,
\label{xi2num}
\end{eqnarray}
where the sum is taken over the whole lattice and $D(R)$ is the Green's $D$ 
function calculated on a lattice of size $L$, which can be written as 
a one-dimensional sum over momentum variables.
For the different values of $c$ we calculate the values of $A_{\rm c}$ and $B_{\rm c}$ 
using~(\ref{xi2num}). Results are given in Table~\ref{tbl:xi2analytic}.

\begin{table}[tb]
\begin{center}
\caption{Values of the $A_{\rm c}$ and $B_{\rm c}$ constants in the scaling of $\xi_2$,
Eq.~(\ref{xi2scaling}).}
\label{tbl:xi2analytic}
\begin{tabular}{|c|c|c|}
\hline
 $c$ & $A_{\rm c}$ & $B_{\rm c}$ \\
\hline
 $\pi/3$   & 1.166... & 0.307... \\
 $\pi/2$   & 0.751... & 0.212... \\
 $2 \pi/3$ & 0.533... & 0.168... \\
\hline
\end{tabular}
\end{center}
\end{table}

\begin{figure}[tb]
\centering
\includegraphics[width=0.44\textwidth]{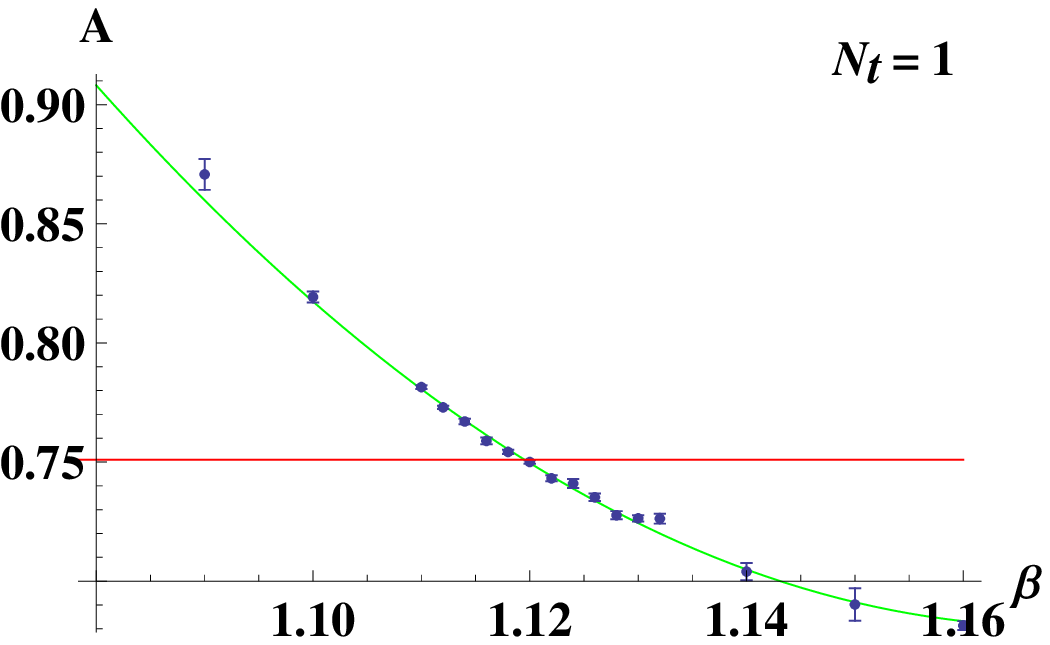}
\hspace{0.2cm}
\includegraphics[width=0.44\textwidth]{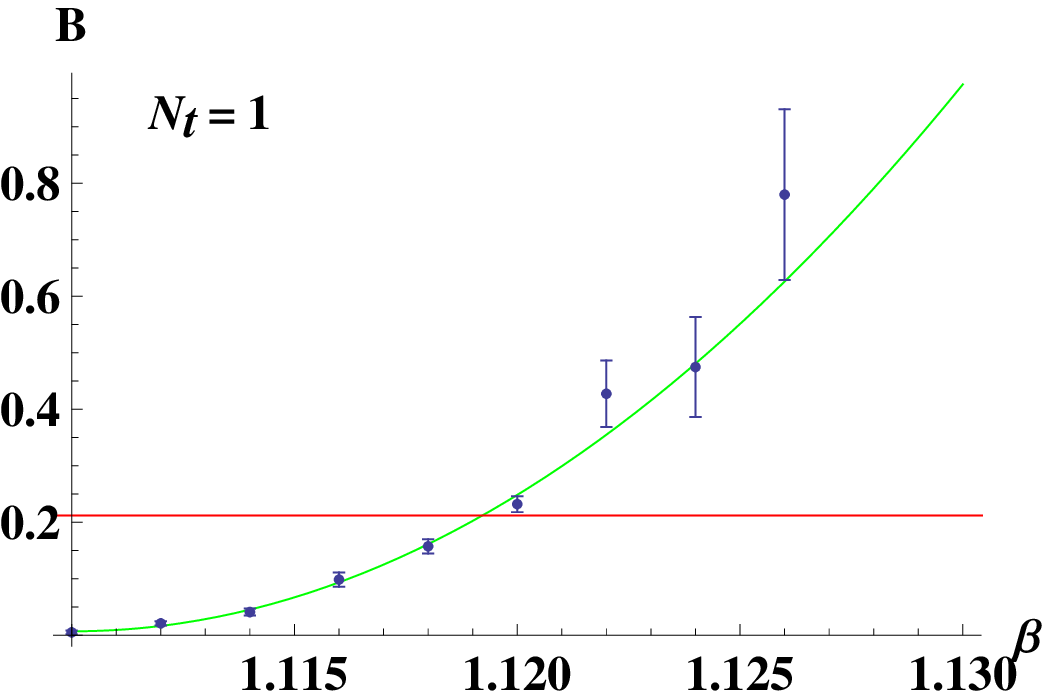}
\includegraphics[width=0.44\textwidth]{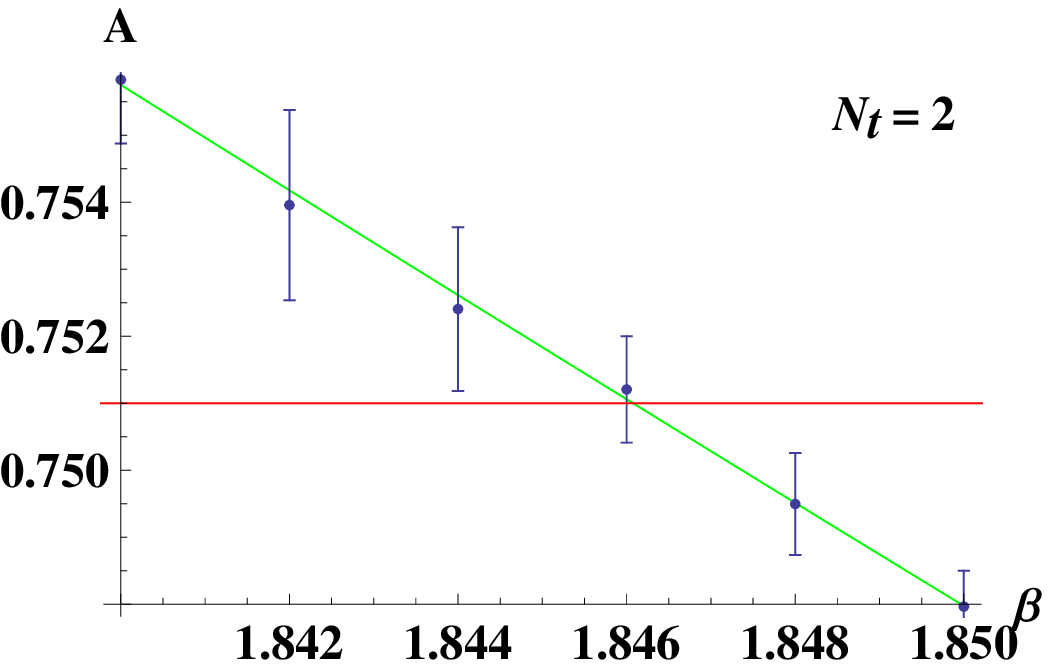}
\hspace{0.2cm}
\includegraphics[width=0.44\textwidth]{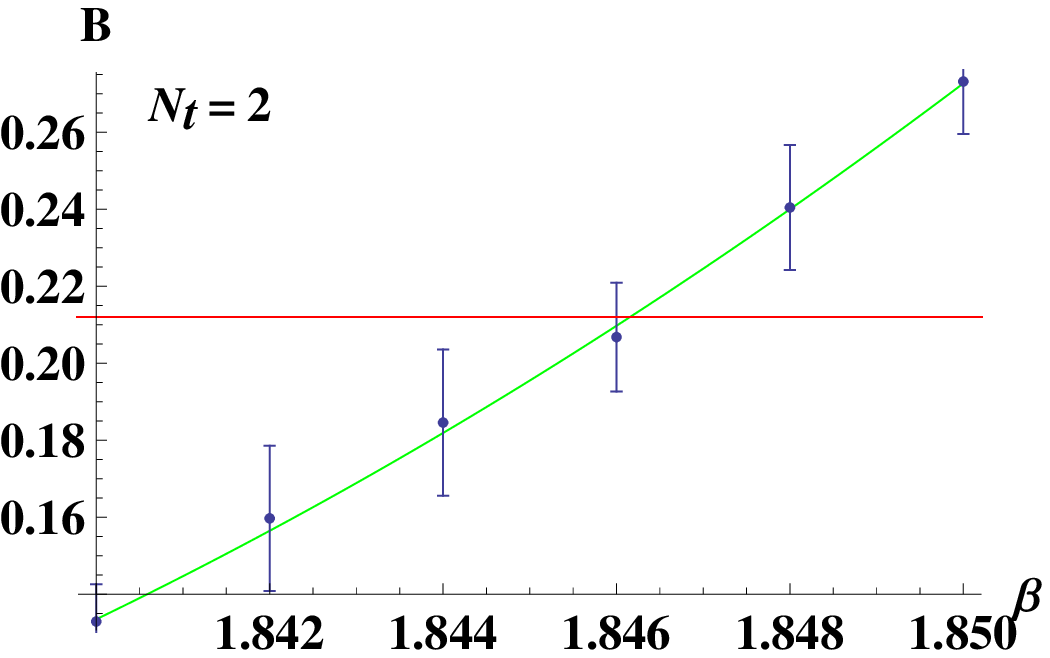}
\includegraphics[width=0.44\textwidth]{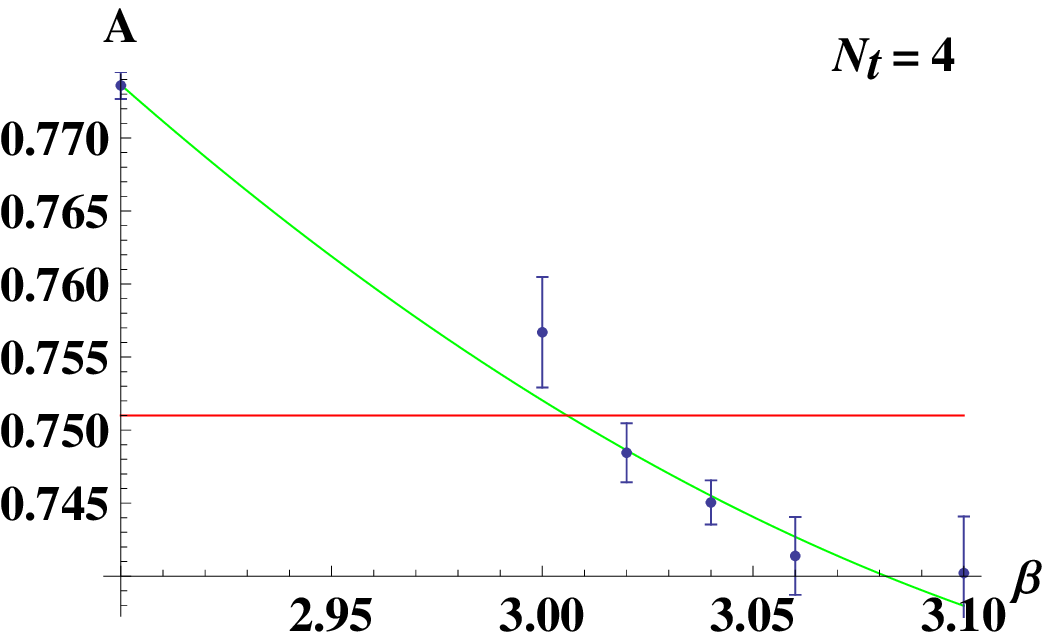}
\hspace{0.2cm}
\includegraphics[width=0.44\textwidth]{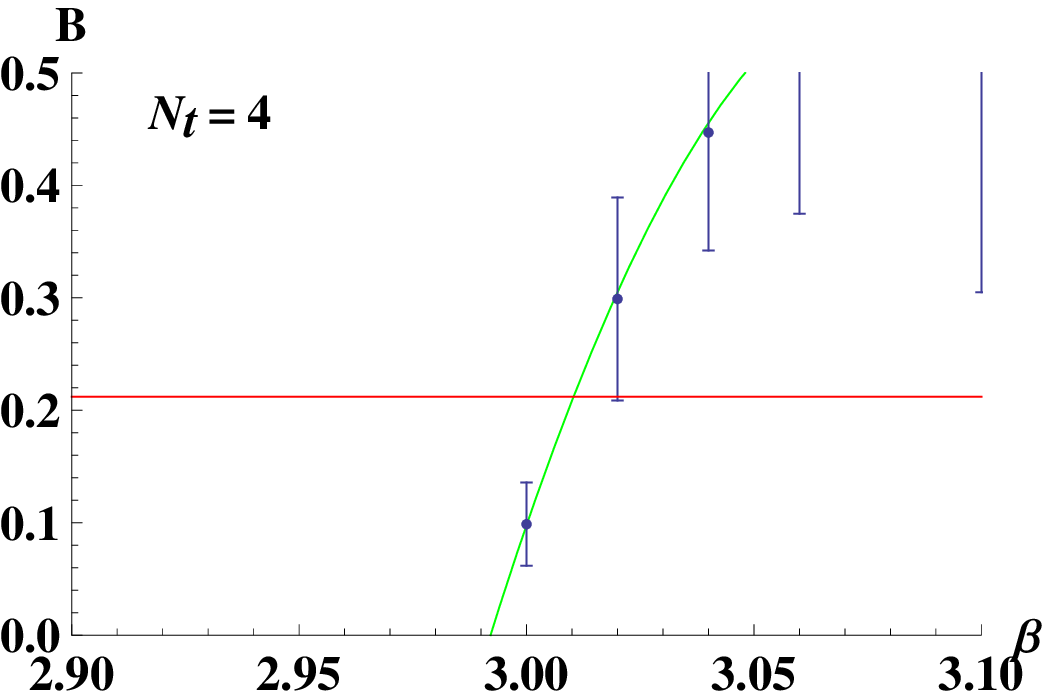}
\includegraphics[width=0.44\textwidth]{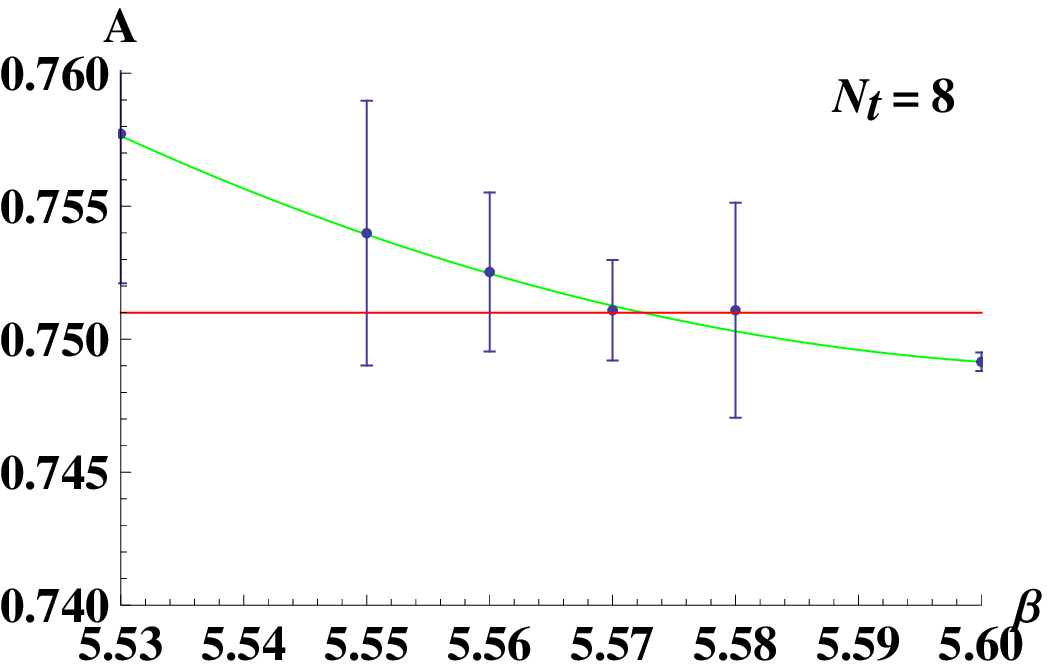}
\hspace{0.2cm}
\includegraphics[width=0.44\textwidth]{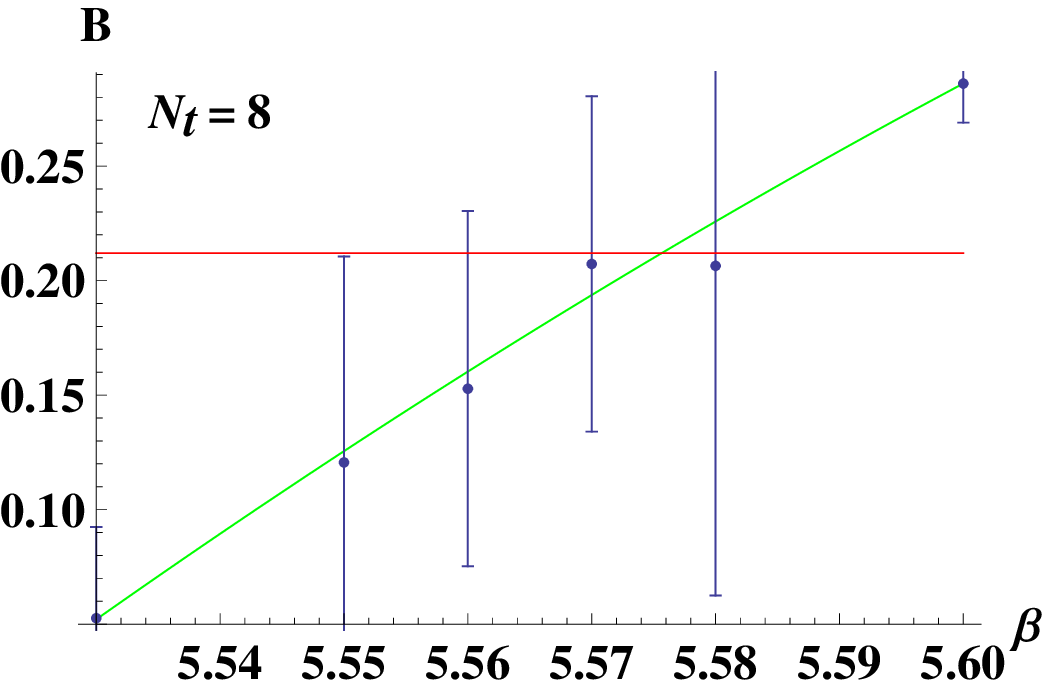}
\caption{Determination of $\beta_{\rm c}$ for $N_t=1$, 2, 4 and 8 
from fits of the scaling~(\ref{xi2scaling_noncrit}) for $A(\beta)$ 
keeping $B(\beta)$ fixed at $B_{\rm c}$ (left panels) and 
for $B(\beta)$ keeping $A(\beta)$ fixed at $A_{\rm c}$ (right panels) in
the case $c = \pi/2$.}
\label{fig:beta_crit}
\end{figure}

To extract the critical point from the scaling of $\xi_2$ with $L$, we use the 
following two methods:
\begin{itemize}
\item At each fixed value of $\beta$, perform the fit to $\xi_2(\beta,L)$ 
with
\begin{equation}
\label{xi2scaling_noncrit}
\frac{\xi_2(\beta,L)}{L} = A(\beta) - \frac{B(\beta)}{\ln L + C(\beta)}\;,
\end{equation}
fixing $B(\beta)$ to the known value $B_{\rm c}$;
$\beta_{\rm c}$ is then found as the point where $A(\beta) = A_{\rm c}$
(see Fig.~\ref{fig:beta_crit}, left panels).
\item The same procedure with $A(\beta)$ fixed to the known value $A_{\rm c}$
and $\beta_{\rm c}$ found as the point where $B(\beta) = B_{\rm c}$
(see Fig.~\ref{fig:beta_crit}, right panels).
\end{itemize}

Results are summarized in Table~\ref{tbl:beta_crit}. We note that the 
values for $\beta_{\rm c}$ obtained for $N_t=8$ are much larger than our
previous determination in the standard formulation of $3D$ 
$U(1)$~\cite{u1_isotropic}, $\beta_{\rm c}=3.06(11)$, thus explaining why we 
found there $\eta\approx 0.49$, a value far away from the expected 1/4.

\begin{table}[tb]
\begin{center}
\caption{Values of $\beta_{\rm c}$ obtained for $c = \pi/3, \pi/2$ and 
$2\pi/3$ for $N_t = 1, 2, 4, 8$, by the two different fit methods described in the text.}
\label{tbl:beta_crit}
\begin{tabular}{|c|c|c|c|}
\hline
 $N_t$ & $c$ & fit $A$, $B$ fixed & fit $B$, $A$ fixed\\
\hline
   & $\pi/3$  & 1.1185(36)  & 1.1184(9) \\
 1 & $\pi/2$  & 1.1195(2)   & 1.1192(2) \\
   & $2\pi/3$ & 1.1176(2)   & 1.1190(6) \\ 
\hline                          
   & $\pi/3$  & 1.84415(15) & 1.84413(10) \\
 2 & $\pi/2$  & 1.8460(8)   & 1.8458(7)  \\
   & $2\pi/3$ & 1.8487(8)   & 1.8488(7) \\
\hline                          
   & $\pi/3$  & 2.991(28)   & 2.991(23)  \\
 4 & $\pi/2$  & 3.005(14)   & 3.010(9)  \\
   & $2\pi/3$ & 3.027(9)    & 3.032(8)  \\
\hline                          
   & $\pi/3$  & 5.567(7)   & 5.565(18)  \\
 8 & $\pi/2$  & 5.572(8)   & 5.573(12)  \\
   & $2\pi/3$ & 5.627(20)  & 5.635(18)  \\
\hline
\end{tabular}
\end{center}
\end{table}

Measuring the correlation function as a function of $R$ in the vicinity of the 
critical points allowed to perform the fit to the function 
$\Gamma(R) = A/2 \exp(- \pi \eta D(R) ) (\ln (R+1)^{2r} + \ln (L-R+1)^{2r}) $.
The expected values for $\eta$ and $r$, given by~(\ref{eta_crit}) 
and~(\ref{index_r}), are summarized in Table~\ref{tbl:crit_indices}.

\begin{table}[tb]
\begin{center}
\caption{Expected critical indices $\eta$ and $r$, defined by~(\ref{eta_crit}) 
and~(\ref{index_r}).}
\label{tbl:crit_indices}
\begin{tabular}{|c|l|l|}
\hline
$c$ & \hspace{1cm} $\eta$ & \hspace{1.5cm} $r$ \\
\hline
$\pi/3$     & 1/9 \ = \ 0.111... & $-$1/36 \ = \ $-$0.028... \\
$\pi/2$     & 1/4 \ = \ 0.25     & $-$1/16 \ = \ $-$0.0625 \\
$2 \pi / 3$ & 4/9 \ = \ 0.444... & $-$1/9 \hspace{0.23cm} = \ $-$0.111... \\
\hline
\end{tabular}
\end{center}
\end{table}             

Since the correlation function values for different $R$ are obtained from the 
same set of measurements, these data cannot be used as independent points in 
a standard fit procedure. Indeed, the procedure we used has been the following:
\begin{itemize}
\item take the correlation function values for $R = 25 - 50$ for $L=256$
and $R = 50 - 100$ for $L = 512$;
\item calculate the covariance matrix $W(R_1,R_2) = 
\left \langle \Gamma(R_1) \Gamma(R_2) \right \rangle - \left\langle \Gamma(R_1) \right\rangle 
\left\langle \Gamma(R_2) \right\rangle$ using the jackknife algorithm;
\item diagonalize $W$, obtaining the eigenvalues $\lambda(R^\prime)$ and the
corresponding transformation matrix $V(R^\prime, R)$ ($R_{\rm min} \leqslant R^\prime \leqslant R_{\rm max}$). 
Since the covariance matrix is symmetric, $V(R,R^\prime)$ is orthogonal,
so that $W(R_1, R_2) = \sum_{R^\prime = R_{\rm min}}^{R_{\rm max}} V(R^\prime, R_1) \lambda(R^\prime) V(R^\prime, R_2)$;
%
%
\item perform a change of basis, obtaining $p(R^\prime) = \sum_{R=R_{\rm min}}^{R_{\rm max}} V(R^\prime, R) \Gamma(R)$.
Since the covariance matrix in the new basis is diagonal, the new $P(R^\prime)$ variables are independent.
\item make a fit to the points $p(R^\prime)$ with weights $1/\lambda(R^\prime)$ (meaning that we minimize 
$\chi^2 = \sum_{R^\prime = R_{\rm min}}^{R_{\rm max}} \left(\Delta p(R^\prime)\right)^2 / \lambda(R^\prime)$).
\end{itemize}
Results are given in Tables~\ref{tbl:cffitd} and~\ref{tbl:cffitd-2}. 

\begin{table}[tb]
\begin{center}
\caption{$\eta$ values obtained from fitting the correlation function 
dependence on $R$ to $\Gamma(R) = A/2^a \exp(-\pi\eta D(R)) [\ln (R + 1)^{2r} 
+ a \ln (L-R+1)^{2r}]$ for $N_t = 1$ and 2; for each $L$, the first line
corresponds to the fit with $r$ fixed at zero, the third to the fit with $a=0$.}
\label{tbl:cffitd}
\small
\begin{tabular}{|c|c|c|c|c|c|}
\hline
 $N_t$, $c$, $\beta_{\rm c}$ & $L$  & $A$ & $\eta$ & $r$ & $\chi^2_{\rm r}$\\
\hline
                   &     & 0.97236(72) & 0.11820(16)  & 0            & 8.51 \\
$N_t=1$            & 256 & 1.0123(36)  & 0.11107(64)  & $-$0.0260(23)  & 1.43 \\
$c=\pi/3$          &     & 0.9602(15)  & 0.1031(19)   & $-$0.0256(31)  & 2.27 \\
\cline{2-6}
$\beta_{\rm c}=1.1195$   &     & 0.96850(57) & 0.11756(11)  & 0            & 5.62 \\
                   & 512 & 0.9859(34)  & 0.11576(36)  & $-$0.0091(18)  & 3.71 \\
                   &     & 0.96698(41) & 0.1000(21)   & $-$0.0348(41)  & 2.28 \\
 \hline
                   &     & 0.9562(16)  & 0.26495(37) & 0            & 7.28 \\
$N_t=1$            & 256 & 1.0532(69)  & 0.2476(13)  & $-$0.0628(44)  & 0.88 \\
$c=\pi/2$          &     & 0.9264(33)  & 0.2256(43)  & $-$0.0670(73)  & 1.64 \\					
\cline{2-6}
$\beta_{\rm c}=1.1195$   &     & 0.9477(13)  & 0.26361(27) & 0            & 5.56 \\
                   & 512 & 1.0035(79)  & 0.25819(77) & $-$0.0287(39)  & 2.74 \\
 		   &     & 0.94626(74) & 0.2145(46)  & $-$0.0983(91)  & 1.66 \\	
 \hline
                   &     & 0.9818(30)  & 0.46843(67) & 0            & 4.77 \\
$N_t=1$            & 256 & 1.176(22)   & 0.4350(40)  & $-$0.119(13)   & 1.35 \\
$c=2\pi/3$         &     & 0.9251(94)  & 0.392(13)   & $-$0.131(22)   & 1.96 \\			
\cline{2-6}
$\beta_{\rm c} = 1.1195$ &     & 0.9651(27)  & 0.46607(54) & 0            & 4.39 \\
                   & 512 & 1.098(20)   & 0.4545(18)  & $-$0.0635(92)  & 2.33 \\
                   &     & 0.9632(19)  & 0.361(14)   & $-$0.210(27)   & 2.02 \\
\hline
                   &     & 0.98785(42) & 0.117439(95) & 0            & 3.03 \\
$N_t=2$            & 256 & 1.0067(32)  & 0.11405(57)  & $-$0.0122(20)  & 1.25 \\
$c=\pi/3$          &     & 0.9812(11)  & 0.1096(12)   & $-$0.0130(22)  & 1.22 \\
\cline{2-6}
$\beta_{\rm c} = 1.8460$ &     & 0.98532(34) & 0.116996(67) & 0            & 2.89 \\
                   & 512 & 0.9971(30)  & 0.11579(31)  & $-$0.0061(15)  & 2.21 \\	
                   &     & 0.98473(38) & 0.1110(21)   & $-$0.0119(42)  & 2.52 \\
 \hline
                   &     & 0.97623(94) & 0.26360(21) & 0            & 2.41 \\
$N_t=2$            & 256 & 1.0193(86)  & 0.2559(15)  & $-$0.0279(55)  & 1.23 \\
$c=\pi/2$          &     & 0.9616(29)  & 0.2455(36)  & $-$0.0306(60)  & 1.19 \\			
\cline{2-6}
$\beta_{\rm c} = 1.8460$ &     & 0.97150(76) & 0.26281(15) & 0            & 2.20 \\
                   & 512 & 0.9951(77)  & 0.26046(77) & $-$0.0122(39)  & 1.88 \\
                   &     & 0.97069(83) & 0.2510(56)  & $-$0.023(11)   & 2.06 \\	
 \hline
                   &     & 0.9704(21)  & 0.46677(47)  & 0            & 2.32 \\
$N_t=2$            & 256 & 1.053(28)   & 0.4523(49)   & $-$0.053(17)   & 1.79 \\
$c=2\pi/3$         &     & 0.9416(92)  & 0.429(12)    & $-$0.064(21)   & 1.70 \\
\cline{2-6}
$\beta_{\rm c} = 1.8460$ &     & 0.9640(16)  & 0.46585(31)  & 0            & 2.09 \\
                   & 512 & 1.024(18)   & 0.4602(16)   & $-$0.0301(86)  & 1.71 \\
                   &     & 0.9626(16)  & 0.428(14)    & $-$0.075(29)   & 1.87 \\		
\hline      
\end{tabular}
\end{center}
\end{table}             
                 
\begin{table}[tb]
\begin{center}
\caption{Same as Table~\ref{tbl:cffitd} for $N_t=4$ and 8.}
\label{tbl:cffitd-2}
\small
\begin{tabular}{|c|c|c|c|c|c|c|}
\hline
 $N_t$, $c$, $\beta_{\rm c}$ & $L$ & $A$ & $\eta$ & $r$ & $\chi^2_{\rm r}$\\  
\hline
                   &     & 1.00035(34) & 0.114548(72) & 0            & 1.70 \\
$N_t=4$            & 256 & 0.9952(41)  & 0.11543(70)  & 0.0033(26)   & 1.66 \\
$c=\pi/3$          &     & 1.0033(12)  & 0.1181(14)   & 0.0061(24)   & 1.40 \\
\cline{2-6}
$\beta_{\rm c} = 3.005$  &     & 0.99975(40) & 0.114423(76) & 0            & 2.19 \\
                   & 512 & 0.9943(28)  & 0.11495(28)  & 0.0028(14)   & 2.07 \\
                   &     & 0.99998(45) & 0.1170(23)   & 0.0051(45)   & 2.18 \\		
\hline
                   &     & 1.00162(84) & 0.25785(18)  & 0            & 1.64 \\
$N_t=4$            & 256 & 0.991(11)   & 0.2596(18)   & 0.0067(69)   & 1.64 \\
$c=\pi/2$          &     & 1.0078(34)  & 0.2656(41)   & 0.0132(69)   & 1.48 \\
\cline{2-6}
$\beta_{\rm c} = 3.005$  &     & 1.0017(12)  & 0.25781(23)  & 0            & 2.97 \\
                   & 512 & 0.9971(93)  & 0.25825(92)  & 0.0023(47)   & 3.02 \\
                   &     & 1.0012(13)  & 0.2521(72)   & $-$0.011(14)   & 3.00 \\		
\hline
                   &     & 1.0045(15)  & 0.45857(31)  & 0            & 1.26 \\
$N_t=4$            & 256 & 0.978(25)   & 0.4629(41)   & 0.017(16)    & 1.25 \\
$c=2\pi/3$         &     & 1.0164(75)  & 0.4738(94)   & 0.026(16)    & 1.18 \\
\cline{2-6}
$\beta_{\rm c} = 3.005$  &     & 1.0025(21)  & 0.45808(39)  & 0            & 2.09 \\
                   & 512 & 0.993(18)   & 0.4589(17)   & 0.0046(88)   & 2.12 \\
                   &     & 1.0023(23)  & 0.455(16)    & $-$0.005(32)   & 2.13 \\
 \hline
                   &     & 0.99993(37) & 0.113792(82) & 0            & 1.83 \\
$N_t=8$            & 256 & 1.0010(45)  & 0.11360(81)  & $-$0.0007(29)  & 1.90 \\
$c=\pi/3$          &     & 0.9989(16)  & 0.1126(17)   & $-$0.0020(29)  & 1.87 \\
\cline{2-6}
$\beta_{\rm c} = 5.572$  &     & 1.00055(34) & 0.113899(65) & 0            & 2.22 \\
                   & 512 & 1.0051(23)  & 0.11339(26)  & $-$0.0024(12)  & 2.08 \\
                   &     & 1.00026(43) & 0.1119(19)   & $-$0.0040(37)  & 2.21 \\
 \hline
                   &     & 1.00008(84) & 0.25608(19)  & 0            & 1.60 \\
$N_t=8$            & 256 & 1.005(11)   & 0.2553(19)   & $-$0.0030(68)  & 1.66 \\
$c=\pi/2$          &     & 0.9971(37)  & 0.2526(42)   & $-$0.0059(71)  & 1.62 \\
\cline{2-6}
$\beta_{\rm c} = 5.572$  &     & 0.99982(81) & 0.25599(15)  & 0            & 1.88 \\
                   & 512 & 1.0090(65)  & 0.25506(67)  & $-$0.0047(33)  & 1.84 \\
                   &     & 0.99974(98) & 0.2552(54)   & $-$0.001(11)   & 1.92 \\		
 \hline
                   &     & 1.0005(13)  & 0.45529(28)  & 0            & 0.97 \\
$N_t=8$            & 256 & 0.995(21)   & 0.4562(35)   & 0.003(13)    & 1.01 \\
$c=2\pi/3$         &     & 1.0003(69)  & 0.4550(84)   & 0.000(14)    & 1.01 \\
\cline{2-6}
$\beta_{\rm c} = 5.572$  &     & 0.9998(21)  & 0.45507(40)  & 0            & 2.21 \\
                   & 512 & 0.997(20)   & 0.4553(19)   & 0.0014(97)   & 2.25 \\
                   &     & 1.008(23)   & 0.470(15)    & 0.030(30)    & 2.20 \\		
\hline
\end{tabular}
\end{center}
\end{table}

Alternative ways to obtain $\eta$ are (i) through the scaling with $L$
of the susceptibility of the magnetization at the critical point, $\chi 
= A L^{-\gamma/\nu}$, using the formula $\eta=2-\gamma/\nu$, or (ii) through 
the scaling with $L$ of the magnetization at the critical point, 
$M = A L^{-\beta/\nu}$, using the formula $\eta=2\beta/\nu$, which assumes the 
validity of the hyperscaling relation $d = 2\beta/\nu + \gamma/\nu = 2$. 
Once we set $c$ in~(\ref{disorder_xy}) to be $2 \pi / K$,
we can build two equally acceptable definitions of the magnetization:
the {\em standard} magnetization, 
\[
M_L= \left\langle \left| M \right| \right\rangle \, , \ \;\;\;\;\; 
M = \sum_x \exp \left( i \frac{c}{N_t} r_x \right) \ ,
\]
and the {\em rotated} one, 
\[
M_R= \left\langle \frac{ {\rm Re} \, M^{K N_t} }{\left| M^{K N_t} \right|} \left| M \right| \right\rangle \ .
\]
It turns out that the hyperscaling relation is better satisfied if
$\beta/\nu$ is extracted from the standard magnetization $M_L$ and
$\gamma/\nu$ from the susceptibility of the rotated magnetization $\chi_{M_R}$, 
with respect to other possible combinations (see Table~\ref{tbl:cmpd} for 
a comparison in some selected cases; here, $d_{M_*,\chi_{M_*}}$, with $*$ equal to 
$L$ or $R$, stands for the value of $d$ obtained by the hyperscaling formula 
$d=2\beta/\nu+\gamma/\nu$ when $\beta/\nu$ is extracted from the scaling of 
the magnetization $M_*$ and $\gamma/\nu$ from that of the magnetization 
susceptibility $\chi_*$).

\begin{table}[tb]
\begin{center}
\caption{Comparison of the results for the hyperscaling relation for 
different ways of obtaining $\beta/\nu$ and $\gamma/\nu$ (see the text
for the explanation of the notation $d_{M_*, \chi_{M_*}}$).}
\label{tbl:cmpd}
\begin{tabular}{|c|c|c|c|c|c|}
\hline
 $N_t$, $c$, $\beta_{\rm c}$ & $L_{\rm min}$ & $d_{M_L, \chi_{M_L}}$ & $d_{M_L, \chi_{M_R}}$ & $d_{M_R, \chi_{M_L}}$ 
& $d_{M_R, \chi_{M_R}}$ \\
\hline
               &  16 & 1.824(20)  & 2.0205(22) & 2.229(34) & 2.425(16) \\
$N_t=1$        &  32 & 1.858(14)  & 2.0169(17) & 2.231(25) & 2.389(13) \\
$c=\pi/2$      &  64 & 1.8848(99) & 2.0137(12) & 2.231(16) & 2.3595(70) \\
$\beta_{\rm c}=1.1195$ & 128 & 1.9058(57) & 2.0127(14) & 2.235(12) & 2.3418(79) \\
               & 192 & 1.9143(41) & 2.0116(22) & 2.234(12) & 2.3312(96) \\
               & 256 & 1.9190(59) & 2.0083(16) & 2.248(19) & 2.337(15)  \\
\hline
               &  16 & 1.9194(84) & 1.9989(12) & 2.500(21) & 2.580(14) \\
$N_t=2$        &  32 & 1.9340(58) & 1.9986(10) & 2.489(16) & 2.554(11) \\
$c=2\pi/3$     &  64 & 1.9449(41) & 1.9989(11) & 2.507(23) & 2.561(20) \\
$\beta_{\rm c}=1.8460$ & 128 & 1.9514(32) & 1.9983(13) & 2.550(32) & 2.597(30) \\
               & 192 & 1.9555(33) & 1.9992(15) & 2.522(50) & 2.566(48) \\
               & 256 & 1.9596(34) & 1.9995(32) & 2.587(63) & 2.627(63) \\
\hline
\end{tabular}
\end{center}
\end{table}   

In Table~\ref{tbl:suspfitmlchimr} we summarize
our determinations of the critical indices $\beta/\nu$, as obtained from
the scaling of the standard magnetization, $\gamma/\nu$, as obtained from
the scaling of the rotated magnetization susceptibility, $\eta=2-\gamma/\nu$, 
as well as $d_{M_L,\chi_{M_R}}$, for $N_t=1,$ 2, 4 and 8.

\begin{table}[ht]
\begin{center}
\caption{Critical indices $\beta/\nu$ (from the standard magnetization), 
$\gamma/\nu$ (from the rotated magnetization susceptibility, $\eta=2-\gamma/\nu$
and $d_{M_L,\chi_{M_R}}$, for $N_t=1,$ 2, 4 and 8.}
\label{tbl:suspfitmlchimr}
\footnotesize
\begin{tabular}{|c|c|c|c|c|c|c|c|c|}
\hline
 $N_t$, $c$, $\beta_{\rm c}$ & $L_{\rm min}$ & $\beta/\nu$ & $\chi^2_{\rm r}$ 
& $\gamma/\nu$ & $\chi^2_{\rm r}$ & $d_{M_L,\chi_{M_R}}$ & $\eta$ \\
\hline
$N_t=1$          & 128 & 0.059304(66) & 2.14 & 1.8919(12)  & 1.32 & 2.0105(13)  & 0.1081(12)  \\
$c=\pi/3$        & 192 & 0.059193(58) & 0.75 & 1.8911(20)  & 1.73 & 2.0095(21)  & 0.1089(20)  \\
$\beta_{\rm c}=1.1195$ & 256 & 0.05914(11)  & 1.07 & 1.8880(11)  & 0.23 & 2.0062(13)  & 0.1120(11)  \\
\hline
$N_t=1$          & 128 & 0.13323(14) & 1.82 & 1.7463(12)  & 1.17 & 2.0127(14)  & 0.2537(12) \\
$c=\pi/2$        & 192 & 0.13299(13) & 0.66 & 1.7456(19)  & 1.58 & 2.0116(22)  & 0.2544(19) \\
$\beta_{\rm c}=1.1195$ & 256 & 0.13288(24) & 0.97 & 1.7425(11)  & 0.21 & 2.0083(16)  & 0.2575(11) \\
\hline
$N_t=1$          & 128 & 0.23609(22) & 1.15 & 1.5461(12)	& 1.11 & 2.0183(16) & 0.4539(12)  \\
$c=2\pi/3$       & 192 & 0.23575(21) & 0.48 & 1.5453(19)	& 1.38 & 2.0168(23) & 0.4547(19)  \\
$\beta_{\rm c}=1.1195$ & 256 & 0.23561(42) & 0.81 & 1.54217(91) & 0.13 & 2.0134(17) & 0.45783(91) \\
\hline
$N_t=2$          & 128 & 0.058718(47) & 1.87 & 1.88167(65) & 0.98 & 1.99910(74) & 0.11833(65) \\
$c=\pi/3$       & 192 & 0.058652(58) & 1.29 & 1.88256(96) & 0.85 & 1.9999(11)  & 0.11744(96) \\
$\beta_{\rm c}=1.8460$ & 256 & 0.05862(12)  & 2.32 & 1.8831(18)  & 1.47 & 2.0003(21)  & 0.1169(18)  \\
\hline
$N_t=2$          & 128 & 0.131917(92) & 0.70  & 1.7357(14)  & 2.77 & 1.9995(15)  & 0.2643(14)  \\
$c=\pi/2$        & 192 & 0.131765(44) & 0.081 & 1.7361(24)  & 4.04 & 1.9996(25)  & 0.2639(24)  \\
$\beta_{\rm c}=1.8460$ & 256 & 0.131805(76) & 0.11  & 1.7390(33)  & 3.47 & 2.0026(35)  & 0.2610(33)  \\
\hline
$N_t=2$         & 128 & 0.23416(20) & 1.88 & 1.53001(93) & 1.37 & 1.9983(13) & 0.46999(93) \\
$c=2\pi/3$      & 192 & 0.23390(27) & 1.51 & 1.53140(97) & 0.69 & 1.9992(15) & 0.46860(97) \\
$\beta_{\rm c}=1.8460$ & 256 & 0.23381(56) & 2.90 & 1.5319(21)  & 1.28 & 1.9995(32) & 0.4681(21)  \\
\hline
$N_t=4$         & 128 & 0.057164(61) & 2.05 & 1.88580(88) & 1.87 & 2.0001(10)  & 0.11420(88) \\
$c=\pi/ 3$      & 192 & 0.05720(11)  & 2.82 & 1.8851(15)  & 2.37 & 1.9995(17)  & 0.1149(15)  \\
$\beta_{\rm c}=3.005$ & 256 & 0.05726(20)  & 4.76 & 1.8832(19)  & 1.59 & 1.9977(23)  & 0.1168(19)  \\
\hline
$N_t=4$         & 128 & 0.12866(10)  & 1.06 & 1.74249(89) &	1.67 & 1.9998(11)  & 0.25751(89) \\
$c=\pi/2$       & 192 & 0.12873(17)  & 1.36 & 1.7417(13)  &	1.84 & 1.9991(17)  & 0.2583(13)  \\
$\beta_{\rm c}=3.005$ & 256 & 0.12884(31)  & 2.19 & 1.7397(11)  &	0.57 & 1.9974(17)  & 0.2603(11)  \\
\hline
$N_t=4$         & 128 & 0.22871(21)  & 1.38 & 1.5423(15)  & 4.02 & 1.9997(19)  & 0.4577(15)  \\
$c=2\pi/3$      & 192 & 0.22880(36)  & 1.94 & 1.5416(25)  & 5.61 & 1.9992(32)  & 0.4584(25)  \\
$\beta_{\rm c}=3.005$ & 256 & 0.22898(69)  & 3.47 & 1.5388(41)  & 6.14 & 1.9967(55)  & 0.4612(41)  \\
\hline
$N_t=8$         & 128 & 0.056945(19)  & 0.24  & 1.88599(56) & 0.82 & 1.99988(60) & 0.11401(56) \\
$c=\pi/3$       & 192 & 0.056935(31)  & 0.32  & 1.88501(56) & 0.34 & 1.99888(62) & 0.11499(56) \\
$\beta_{\rm c}=5.572$ & 256 & 0.056882(22)  & 0.063 & 1.88589(53) & 0.12 & 1.99965(57) & 0.11411(53) \\
\hline
$N_t=8$         & 128 & 0.128167(80) & 0.56     & 1.74284(72) & 0.85 & 1.99917(88) & 0.25716(72) \\
$c=\pi/2$       & 192 & 0.12820(14)  & 0.79     & 1.74186(80) & 0.52 & 1.9983(11)  & 0.25814(80) \\
$\beta_{\rm c}=5.572$ & 256 & 0.1279502(15)& 3.3 10$^{-5}$ & 1.7410(16)  & 0.71 & 1.9969(16)  & 0.2590(16)  \\
\hline
$N_t=8$         & 128 & 0.227822(65) & 0.17  & 1.54418(53) & 0.46  & 1.99982(66) & 0.45582(53) \\
$c=2\pi/3$      & 192 & 0.22778(11)  & 0.23  & 1.54502(31) & 0.083 & 2.00059(53) & 0.45498(31) \\
$\beta_{\rm c}=5.572$ & 256 & 0.227607(50) & 0.021 & 1.54541(58) & 0.097 & 2.00063(68) & 0.45459(58) \\
\hline
\end{tabular}
\end{center}
\end{table}

\begin{figure}[tb]
\centering
\includegraphics[width=0.7\textwidth]{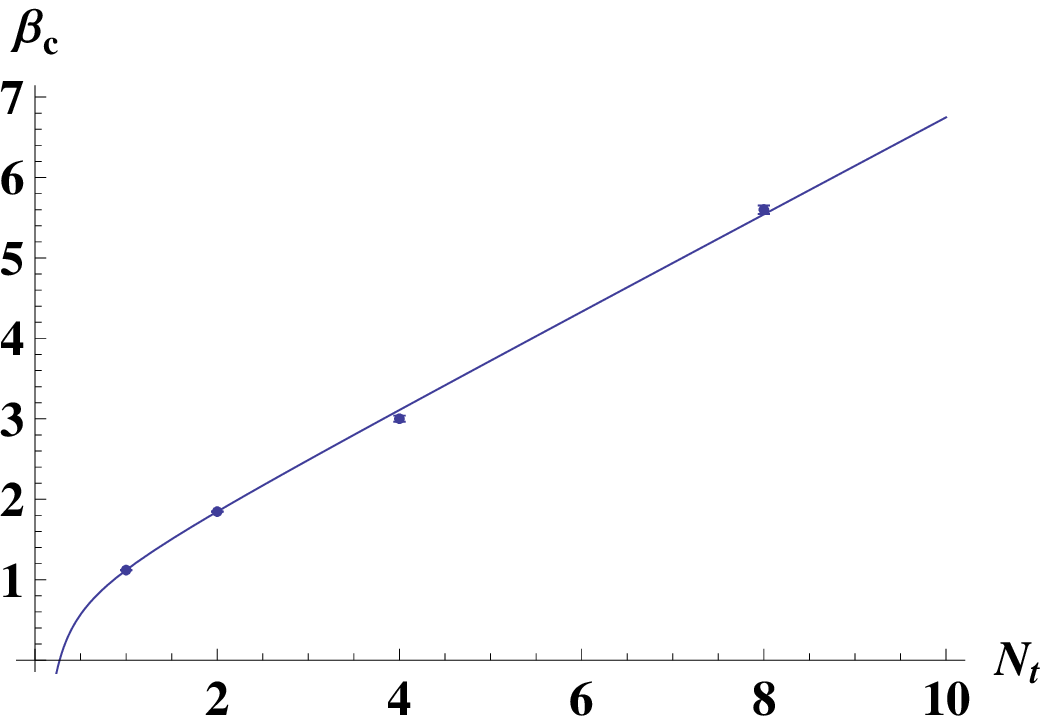}
\caption{Fitting curve for the dependence on $N_t$ of the critical couplings.} 
\label{fig:clim}
\end{figure}

Finally, we constructed the continuum limit fitting the critical couplings 
$\beta_{t,\rm c}$ from Table~\ref{tbl:beta_crit} using several dependences on 
$N_t$. As estimate of the critical point we took the half-sum of the 
largest and of the smallest values obtained, for a given $N_t$, considering
the three possible choices for $c$ and the two fitting methods; as estimate
of its uncertainty, we took the half-difference of the same values.
The best fit is given by the function $\beta_{t,\rm c} = 0.772(90) 
+ 0.600(29) N_t - 0.252(64) / N_t, \ \chi^2 = 9.13$ (see Fig.~\ref{fig:clim}). 

\section{Summary}

In this paper we have studied the disorder operator~(\ref{disorder_u1}) in the 
dual formulation of the finite-temperature $3D$ $U(1)$ LGT.  
We obtained and analyzed the RG equations in the Villain formulation of the 
model. 
These equations describe the critical behavior across the deconfinement phase 
transition. 
The Wilson formulation in its dual representation has been studied via 
numerical simulations for three values of the constant $c$ entering the 
definition~(\ref{disorder_u1}). 
Our main findings can be shortly summarized as follows. 

\begin{itemize} 

\item We have calculated analytically the critical indices $\eta$ and $r$ for 
any values of $c$, using the conventional RG. We have found that the critical 
behavior of the disorder operator of the $3D$ $U(1)$ gauge theory is governed 
by the critical behavior of the corresponding operator in the $2D$ $XY$ model. 
It is important to stress that the index $r$, describing the leading 
logarithmic correction, is negative for the disorder operator. 

\item Using a cluster algorithm, we have simulated the dual form of the $3D$ 
$U(1)$ LGT, computed the disorder operator, the second moment correlation 
length, the standard magnetization and the rotated magnetization of the dual 
variables for three values of the constant $c=\pi/3$, $\pi/2$ and $2\pi/3$. 
In this way we have located critical points of the finite-temperature model 
for $N_t=1$, 2, 4 and 8, computed the critical indices $\eta$ and $r$ and 
checked the hyperscaling relation. 

\item We have computed the critical points for several temporal extensions 
$N_t$. In the continuum limit we have found $T_{\rm c} \approx 0.600 g^2$. This value 
agrees with the value obtained in~\cite{twist_rg} from the study of the 
critical behavior of the twist free energy. 

\end{itemize} 

It is important to stress that, while the index $\eta$ agrees reasonably with 
analytical and universality predictions for all values of $c$, this is not always 
the case for the index $r$. 
However, we would like to stress that both RG study and numerical simulations 
show that this index is negative. This is an interesting property of the 
disorder operator.

Our final remark concerns the check of the hyperscaling relation. 
Table~\ref{tbl:cmpd} shows the value of $d$ extracted in four possible ways. 
One can conclude that the hyperscaling relation is satisfied only when 
$\beta/\nu$ is calculated from the conventional magnetization and
$\gamma/\nu$ from the susceptibility of the rotated magnetization. 
This remains true for all values of $c$ and might indicate some important 
property of the disorder operator which we miss at the moment. 

Summarizing, we would like to stress that this work, together with our previous 
studies, leaves little doubt, if any, that the deconfinement phase transition 
in finite-temperature $3D$ $U(1)$ LGT belongs to the universality class 
of the $2D$ $XY$ model, thus supporting the Svetitsky-Yaffe conjecture.


\begin{thebibliography}{99} 

%
\bibitem{svetitsky} B.~Svetitsky, L.~Yaffe, Nucl. Phys. {\bf B210} (1982) 423.
%
\bibitem{berezin} V.~Berezinskii, Sov. Phys. JETP {\bf 32} (1971) 493. 
%
\bibitem{kosterlitz1} J.~Kosterlitz, D.~Thouless, J. Phys. {\bf C6} (1973) 1181;
J.~Kosterlitz, J. Phys. {\bf C7} (1974) 1046.
%
\bibitem{mwtheorem} N.~Mermin, H.~Wagner, Phys. Rev. Lett. {\bf 22} (1966) 1133.
%
\bibitem{parga} N.~Parga, Phys. Lett. {\bf B107} (1981) 442. 
%
\bibitem{borisenko} O.~Borisenko, PoS LATTICE {\bf 2007} (2007) 170.
%
\bibitem{twist_rg} O.~Borisenko, V.~Chelnokov, Phys. Lett. {\bf B730} (2014) 
226. 
%
\bibitem{3dzn_strcoupl} O.~Borisenko, V.~Chelnokov,  G.~Cortese, R.~Fiore, 
M.~Gravina, A.~Papa, I.~Surzhikov, Phys. Rev. {\bf E86} (2012) 051131. 
%
\bibitem{3dzn_isotropic1} O.~Borisenko, V.~Chelnokov,  G.~Cortese, M.~Gravina, 
A.~Papa, I.~Surzhikov, Nucl. Phys. {\bf B870} (2013) 159.
%
\bibitem{3dzn_isotropic2} O.~Borisenko, V.~Chelnokov, M.~Gravina, A.~Papa, 
Nucl. Phys. {\bf B888} (2014) 52.
%
\bibitem{pos_lat} O.~Borisenko, V.~Chelnokov,  G.~Cortese, R.~Fiore, M.~Gravina,
A.~Papa, I.~Surzhikov, PoS LATTICE {\bf 2012} (2012) 270; 
O.~Borisenko, V.~Chelnokov,  G.~Cortese, M.~Gravina, A.~Papa, I.~Surzhikov, 
PoS LATTICE {\bf 2013} (2013) 347. 
%
\bibitem{mcfinitet} P.~Coddington, A.~Hey, A.~Middleton, J.~Townsend, 
Phys. Lett. {\bf B175} (1986) 64.
%
\bibitem{beta_szero} O.~Borisenko, M.~Gravina, A.~Papa, J. Stat. Mech. 
{\bf 0808} (2008) P08009. 
%
\bibitem{u1_isotropic} O.~Borisenko, R.~Fiore, M.~Gravina, A.~Papa, 
J. Stat. Mech. {\bf 1004} (2010) P04015.
%
\bibitem{xy_disorder} J.~Fr\"ohlich, T.~Spencer, Commun. Math. Phys. {\bf 81} 
(1981) 527.
%
\bibitem{nelson} D.~R.~Nelson, J.~Kosterlitz, Phys. Rev. Lett. {\bf 39} (1977) 
1201.
%
\bibitem{itzykson} C.~Itzykson, J.-M.~Drouffe, {\it Statistical field theory}, 
Volume 1, 
{\it From Brownian motion to renormalization and lattice gauge theory},  
Cambridge University Press, 1989. 
%
\bibitem{hasenbusch} M.~Hasenbusch, J. Phys. {\bf A38} (2005) 5869. 
%
\bibitem{binder_xy} M.~Hasenbusch, arXiv:0804.1880 [cond-mat].
%
\bibitem{dual_cluster} M.~Hasenbusch, G.~Lana, M.~Marcu and K.~Pinn, 
Phys. Rev. {\bf B46} (1992) 10472.

\end{thebibliography}
\end{document}